\documentclass[twocolumn,twoside,slac_two]{revtex4}
\usepackage{graphicx}
\usepackage{fancyhdr}
\usepackage{graphics}
\usepackage{epstopdf}
\usepackage{textpos}
\begin{document}

\title{\centering Solar Neutrinos in 2011}


\author{
\centering
\begin{center}
Alvaro E. Chavarria on behalf of the Borexino Collaboration
\end{center}}
\affiliation{\centering Princeton University Physics Department, New Jersey, 08544, United States}
\begin{abstract}
I give an overview of the recent developments in the solar neutrino field. I focus on the Borexino detector, which has uncovered the solar neutrino spectrum below 5\,MeV, providing new tests and confirmation for solar neutrino oscillations. I report on the updated measurements of the $^8$B solar neutrino flux by water Cherenkov and organic scintillator detectors. I review the precision measurement of the $^7$Be solar neutrino flux by Borexino and the search for its day-night asymmetry. I present Borexino's latest result on the study of pep and CNO neutrinos. Finally, I discuss the outstanding questions in the field and future solar neutrino experiments.
\end{abstract}

\maketitle
\thispagestyle{fancy}


\section{Prelude \label{sec:prelude}}

In 2002, results from the SNO collaboration solved the long-standing solar neutrino problem by demonstrating that solar neutrinos oscillate, i.e. the flavor content of solar neutrinos detected on Earth is different from their initial electron neutrino state when produced in the Sun~\cite{snosnp}. Further measurements of the $^{8}$B solar neutrino flux by SNO~\cite{snolatest} and Super-Kamiokande~\cite{sklatest}, as well as the results on reactor anti-neutrino oscillations reported by KamLAND~\cite{kamlatest}, have allowed to measure the two relevant free parameters of the established oscillation model, MSW-LMA~\cite{msw}. The latest experimental values for these parameters are $\theta_{12}=0.45\pm0.03$ and $\Delta m^{2}_{21}=(7.4\pm0.2)\times 10^{-5}$\,eV$^2$~\cite{snolatest}.

\begin{figure}
\includegraphics[width=\columnwidth]{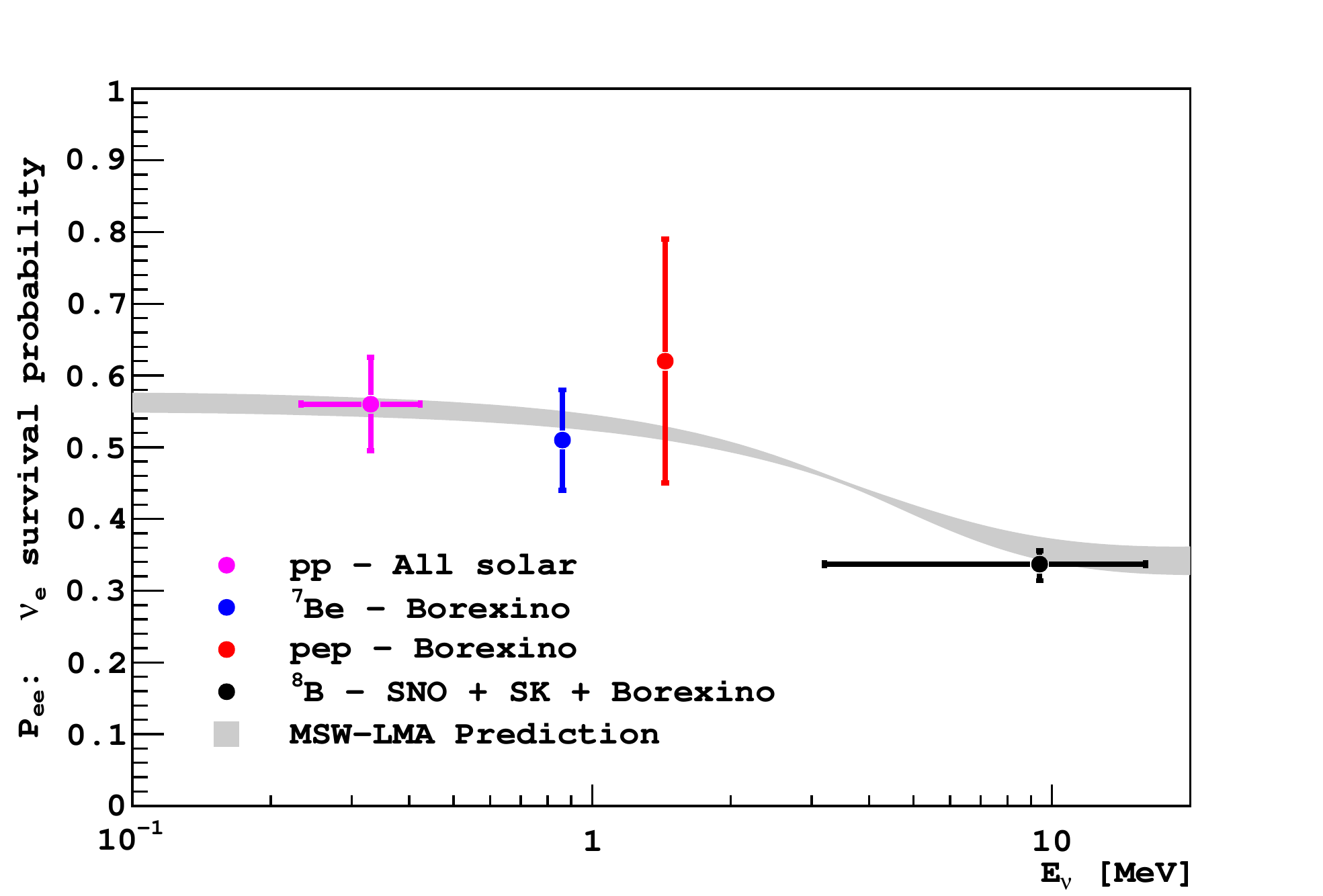}
\caption{ Probability, as a function of neutrino energy, that an electron neutrino produced in the Sun will be detected as an electron neutrino on Earth. The prediction of MSW-LMA is shown by the gray band. The higher survival probability region at low energies is where vacuum-dominated oscillations occur. As the neutrino energy increases, matter effects become important and hence the lower survival probability region at high energies are due to matter-enhanced oscillations. In between there is the MSW transition region.
The markers on the plot correspond to solar neutrino flux measurements performed by various experiments.}
\label{fig:pee}
\end{figure}

Even though the precise study of $^8$B neutrinos allowed for the determination of the oscillation parameters, key features of the oscillation model remained to be observed. In particular, the predicted vacuum-dominated oscillation of sub-MeV neutrinos and the corresponding transition to matter-enhanced oscillations in the 1--3\,MeV region (see Figure~\ref{fig:pee}). In the last five years, experimental efforts in the solar neutrino field have focused on detecting neutrinos with ever decreasing energy, not only to confirm these details of MSW-LMA but also to verify the neutrino fluxes predicted by the Standard Solar Models (SSMs).\\

\section{Solar Neutrinos \label{sec:solar-nu}}

\begin{figure}
\includegraphics[width=\columnwidth]{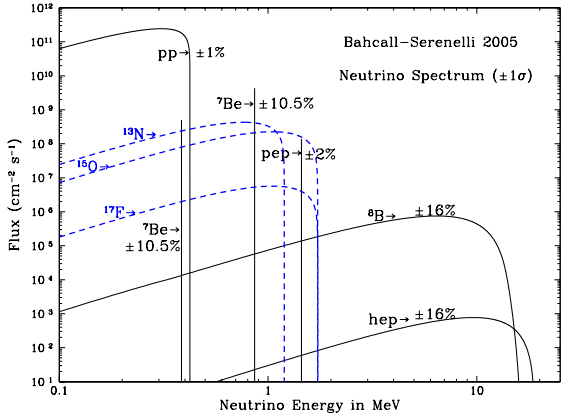}
\caption{Energy spectra and fluxes at Earth for solar neutrinos predicted by the Standard Solar Model (SSM). Black lines correspond to species from the pp fusion chain, while blue lines correspond to species from the sub-dominant CNO cycle. Graph taken from~\cite{bahcall}.}
\label{fig:snu_spec_flux}
\end{figure}

Figure~\ref{fig:snu_spec_flux} shows the energy spectra and fluxes at Earth for solar neutrinos predicted by the SSM~\cite{ssm2011}. The Sun is powered by the fusion of four protons into a $^{4}$He nucleus, with the production of two neutrinos. The pp chain and the CNO cycle are the two main mechanisms for this process. The pp chain accounts for $\sim$99\% of the energy production in the Sun and, therefore, produces the largest flux of neutrinos.

The pp and pep fluxes are strongly correlated and their values are predicted with highest precision by relying on the solar luminosity constraint~\cite{tenthousand}. The uncertainties in the other fluxes from the pp chain ($^7$Be, $^8$B and hep) are larger due to uncertainties in nuclear cross-sections and the solar opacity~\cite{tenthousand}. The fluxes of neutrinos from the CNO cycle are less precisely predicted because of conflicting estimates of the abundances of elements heavier than helium, which have led to the solar composition (metallicity) problem~\cite{metallicity}. Therefore, measurements of the multiple solar neutrino species are crucial for the understanding of solar composition and dynamics.

Due to their low photon yield, water Cherenkov experiments~\cite{snodetectorpaper, skdetectorpaper} can only detect $^8$B and hep neutrinos (3.5\,MeV energy threshold), while radiochemical experiments~\cite{cldetector, sagedetector} can only measure the integrated count rate of neutrinos above the charge-current interaction threshold (0.81\,MeV for $^{37}$Cl and 0.23\,MeV for $^{71}$Ga), and therefore unable to distinguish the different neutrino species. Therefore, in order to measure the particular low energy neutrino fluxes and uncover most of the solar neutrino spectrum, organic liquid scintillator detectors capable of performing low energy spectroscopy have been developed.

\section{The Low-Energy Frontier \label{sec:low-energy-frontier}}

Organic liquid scintillator detectors~\cite{bxdetectorpaper, kldetectorpaper} have been used due to their high light yield ($\sim$$10^4$\,photons per MeV of deposited energy) and the feasibility to produce large, unsegmented target masses (hundreds of tons or more) of high-purity material. Below 5\,MeV, backgrounds from decays from natural radioactivity are significant, with an intensity in rock of $\sim$1 decay per second per gram. The expected rate of solar neutrino interactions are at least 10 orders of magnitude lower than this and, therefore, extreme levels of radio-purity are required in the target. Furthermore, highly-penetrating $\gamma$-rays from the peripheral structure holding the target material can contribute a significant count rate, regardless of the target's radio-purity, making it necessary for the detector to be unsegmented and capable of fiducialization.

\begin{figure}
\includegraphics[width=\columnwidth]{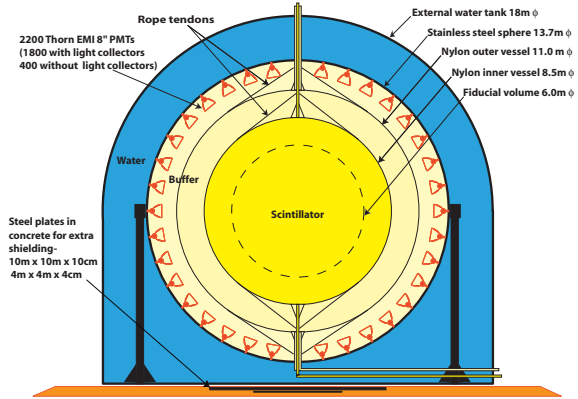}
\caption{The Borexino detector~\cite{bxdetectorpaper}. It was designed to detect sub-MeV solar neutrinos. It features a high light-yield, ultra-pure liquid scintillator target. A non-scintillating buffer region serves as shielding for external $\gamma$-rays. Its location at a deep underground site and its muon veto suppress cosmic backgrounds.}
\label{fig:bx_sketch}
\end{figure}

Figure~\ref{fig:bx_sketch} is a schematic of Borexino~\cite{bxdetectorpaper}, the epitome of low-background detectors, which is located at the Gran Sasso National Laboratories (LNGS).  It features an active target of 278\,tons of ultra-pure pseudocumene-based scintillator contained in a nylon vessel, which is surrounded by 898\,tons of highly-pure pseudocumene-based non-scintillating buffer fluid that serves as shielding from $\gamma$-ray background from the peripheral structure. A second nylon vessel separates the buffer into two segments and prevents radon diffusion toward the inner target. To achieve the low-radioactivity goals, great care was taken in the fluid purification~\cite{purification} and in the construction of the nylon vessels~\cite{nylonvessels}. The laboratory site offers an overburden of 3800\,m water-equivalent and an instrumented water tank surrounding the detector serves as a veto for the residual cosmic muon flux.

Borexino sees the electron recoils from neutrino-electron elastic scattering in the active target by measuring the produced scintillation light with photomultiplier tubes. The energy of the event can be reconstructed from the detected number of photoelectrons ($\sim$500\,photoelectrons per MeV), while the position of the event can be calculated from the photoelectron detection times, considering that the scintillation light is emitted isotropically~\cite{tofalg}. 

\begin{figure}
\includegraphics[width=\columnwidth]{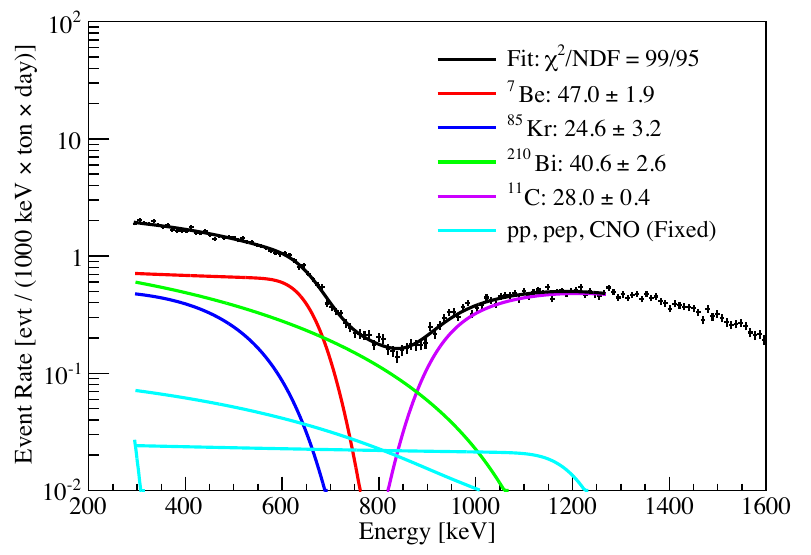}
\caption{The energy spectrum of electron and positron recoils in Borexino's fiducial volume. The colored lines present the best-fit result for the species considered in the extraction of the $^7$Be neutrino interaction rate. $^{85}$Kr and $^{210}$Bi decays are due to contamination present in the scintillator target, while $^{11}$C is produced in situ by cosmic muons. The spectral feature of the box-like structure of electron recoils from $^7$Be neutrinos is evident.}
\label{fig:be_spectrum}
\end{figure}

Figure~\ref{fig:be_spectrum} shows the electron recoil spectrum in Borexino's 75\,ton fiducial volume after all analysis cuts. The count rate above the low-energy background due to $^{14}$C, i.e. $>$0.3\,MeV, is $10^{-5}$\,counts per second per ton, sufficiently low for the successful identification of the electron recoil spectra from solar neutrinos.

\section{Low-threshold $^{8}$B neutrino measurements}

\begin{figure*}[t]
\centering
\includegraphics[width=135mm]{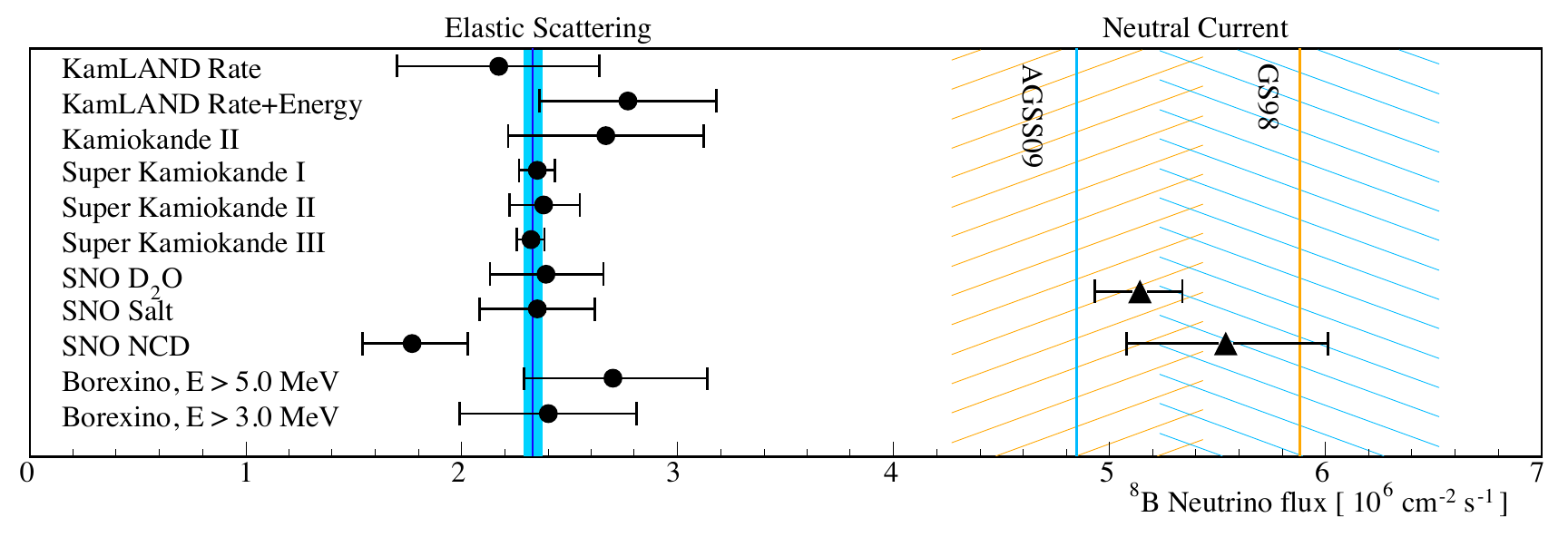}
\caption{Summary of the measured $^8$B solar neutrino flux by different experiments. All experiments can provide a value for the flux from the measured rate of neutrino-electron elastic scattering interactions, assuming no neutrino oscillations. SNO provides a flavor-independent value of the total flux through the measured rate of neutral current interactions. The line-shaded regions represent predictions by SSMs that assume different solar composition. The lower estimated flux from the rate of elastic scattering interactions, evidence of solar neutrino oscillations, is present in all experimental data. Figure taken from~\cite{klb8}.}
\label{fig:b8_le}
\end{figure*}

In the past two years SNO~\cite{snolatest}, Super-Kamiokande~\cite{sklatest}, Borexino~\cite{bxb8} and KamLAND~\cite{klb8} have performed updated measurements of the $^{8}$B solar neutrino flux (see Figure~\ref{fig:b8_le}). The Borexino and SNO results have a lower electron energy threshold at 3.0\,MeV and 3.5\,MeV, respectively, increasing the energy range of electron recoils considered for the measurement of the $^{8}$B neutrino survival probability. Due to the decreasing flux of $^{8}$B neutrinos below 10\,MeV and the decreasing interaction cross-sections toward low energies, these results do not provide any significant information about the survival probability of solar neutrinos below 5\,MeV. Nevertheless, the results are consistent with expectations and have overcome the technical challenge of measuring the feeble $^{8}$B neutrino signal (a fraction of an interaction per day in Borexino's fiducial volume) in an energy region where radioactive background contributions can be significant. The latest, combined result of the $^{8}$B neutrino survival probability is shown in Figure~\ref{fig:pee}.

\section{Precision measurement of $^{7}$Be neutrinos}

Following up on previous results~\cite{bxbe7old}, Borexino has released a measurement of the monoenergetic $^{7}$Be solar neutrino interaction rate with 5\% uncertainty~\cite{bxbe7new}. This result was made possible by an extensive calibration campaign with internal and external radioactive sources, which allowed for a good understanding of the detector's energy response, the spatial reconstruction of the electron recoils and the efficiency of the analysis cuts. The 0.86\,MeV $^7$Be neutrino interaction rate was extracted from a fit to the energy spectrum (see Figure~\ref{fig:be_spectrum}) and is 46.0$\pm$1.5(stat)$^{+1.5}_{-1.6}$(syst) interactions per day per 100\,tons. Assuming the predicted $^7$Be neutrino flux by the SSM, this result is 5$\sigma$ lower than would be expected in the absence of solar neutrino oscillations. If the electron neutrinos are oscillating to $\mu$ or $\tau$ neutrinos, this corresponds to an electron neutrino survival probability of 0.51$\pm$0.07 at 0.86\,MeV. This result, combined with the Gallium radiochemical~\cite{sagedetector} and $^{8}$B neutrino measurements~\cite{snolatest, sklatest}, allows to infer the survival probability for pp neutrinos. The results are summarized in Figure~\ref{fig:pee} and represent the observation of vacuum-dominated oscillations of sub-MeV neutrinos, as predicted by MSW-LMA.

Borexino has also published a result on the absence of a difference between the day and night  $^{7}$Be neutrino interaction rates~\cite{daynight}. The rate difference, expressed as a fraction of the average interaction rate, is 0.001$\pm$0.012(stat)$\pm$0.007(syst). This is the strongest constraint on matter effects for sub-MeV neutrinos and is further confirmation of MSW-LMA. In particular, ``with this result the LOW region of MSW parameter space is, for the first time, strongly disfavored by solar neutrino data alone"~\cite{daynight}.

\section{First direct study of pep and CNO neutrinos}

The usefulness of $^7$Be neutrinos to test neutrino oscillations is limited by the uncertainty in the predicted flux by the SSM, which is 7\%~\cite{ssm2011}, comparable to the uncertainty achieved in the flux measurement by Borexino. The monoenergetic 1.44\,MeV pep neutrino lends itself to more stringent tests on solar neutrino oscillations, as the uncertainty in the SSM predicted flux is $<$2\% and its energy is within the transition region of the electron neutrino survival probability  (see Figure~\ref{fig:pee}), which is particularly sensitive to new physics~\cite{newphys}.

The interactions from neutrinos from the solar CNO cycle have energies and fluxes which are predicted to be comparable to those from pep neutrinos (see Figures~\ref{fig:snu_spec_flux}). The predicted fluxes of CNO neutrinos are the ones with the largest uncertainty in the SSM~\cite{ssm2011} and their determination may offer key information for the resolution of the solar composition problem~\cite{solarnucomp}. Thus, the measurement of the pep and CNO neutrino interaction rates is of crucial importance for the understanding of neutrino oscillations and solar astrophysics.

These interaction rates are predicted to be ten times lower than the $^7$Be neutrino interaction rate (see Figure~\ref{fig:be_spectrum}), demanding more statistics and a better understanding of detector backgrounds for their measurement.

\begin{figure}
\includegraphics[width=\columnwidth]{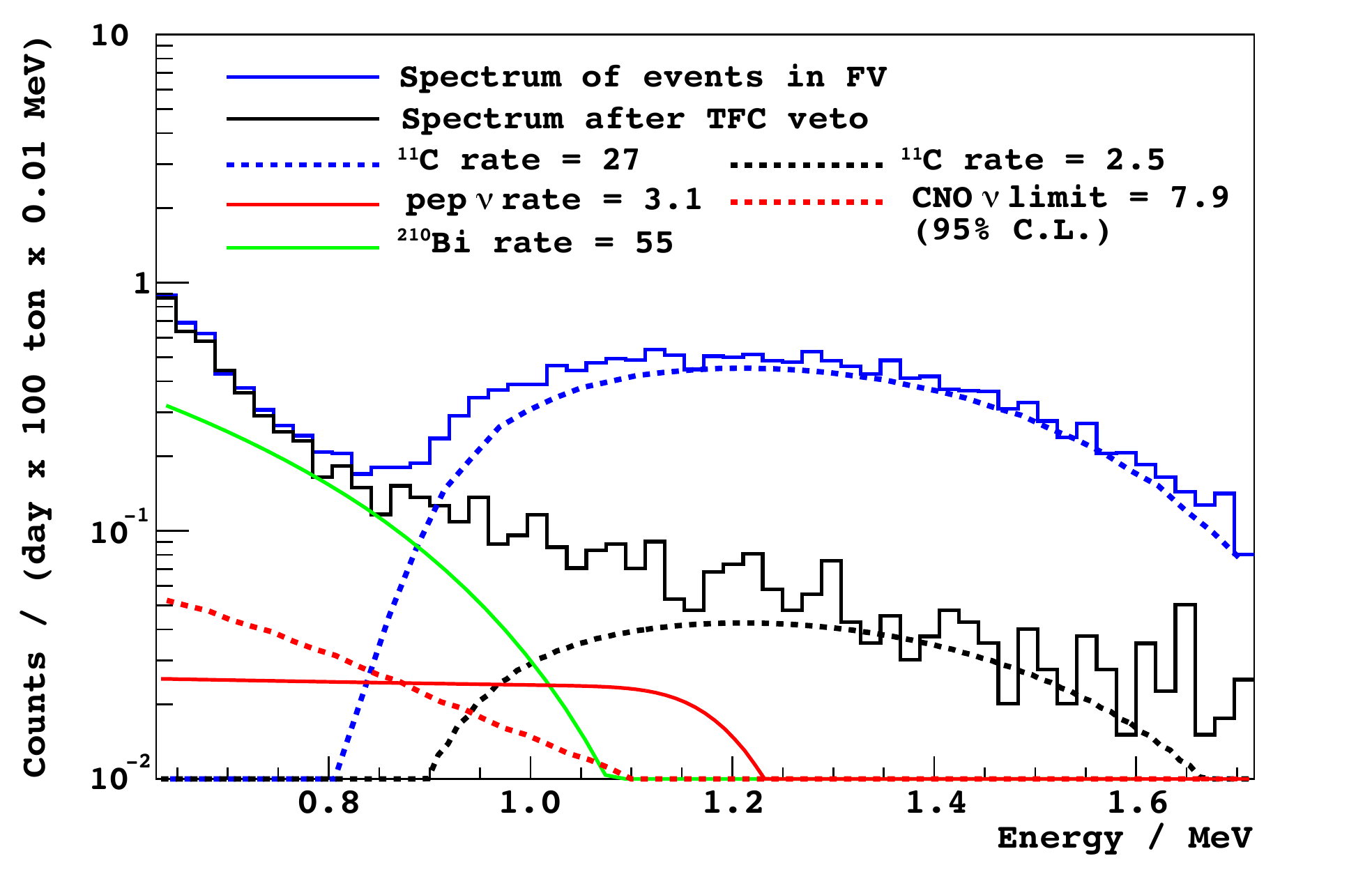}
\caption{Energy spectrum in Borexino's fiducial volume before (blue) and after (black) the background rate of $^{11}$C decays is decreased significantly by relying on its space and time correlation with cosmogenic neutrons. The faint electron recoil signal from pep and CNO neutrinos has then been extracted by a multivariate fit. $^{210}$Bi is now the dominant background in this energy region.}
\label{fig:pep_cno}
\end{figure}

Borexino has released the first measurement of the pep neutrino flux and the strongest constraint on the flux of neutrinos from the CNO cycle~\cite{pep}. The main challenge in this study was the suppression of the background due to decays of the cosmogenic $\beta^+$ emitter $^{11}$C, whose rate in Borexino is an order of magnitude larger than the signal from pep and CNO neutrinos (see Figure~\ref{fig:be_spectrum}). As free neutrons are produced in the scintillator when cosmic muons interact with $^{12}$C nuclei to produce $^{11}$C~\cite{c11cris}, it is possible to take advantage of the space and time correlation of the neutron capture $\gamma$-rays and the $^{11}$C decays to substantially decrease the $^{11}$C background (see Figure~\ref{fig:pep_cno}). Furthermore, by considering the differences in the time profiles of the emitted scintillation light between electron and positron recoils (mostly due to the formation of ortho-positronium in the scintillator before annihilation~\cite{annihilation}), Borexino has been able to perform a multivariate fit to extract a pep rate of 3.1$\pm$0.6(stat)$\pm$0.3(syst) interactions per day per 100\,tons. The absence of the pep signal is disfavored at 98\% C.L. The measured pep interaction rate corresponds to an electron neutrino survival probability of 0.62$\pm$0.17 at 1.44\,MeV (see Figure~\ref{fig:pee}). Assuming the pep flux predicted by the SSM, Borexino has also provided a 95\% C.L. upper limit on the flux of neutrinos from the CNO cycle of $<7.9$ interactions per day per 100\,tons. These results are consistent with MSW-LMA and the SSM predictions.

\section{The future}

Borexino has taken the first steps toward the measurement of pep and CNO solar neutrinos but the results do not yet have sufficient precision to test the MSW transition region or to resolve the solar composition problem. In order to improve the result, Borexino has undertaken since July 2010 a purification campaign to decrease the radioactive backgrounds, in particular the $^{210}$Pb present in the scintillator, which is the source of $^{210}$Bi (see Figure~\ref{fig:pep_cno}). The campaign has been successful in reducing the levels of $^{85}$Kr, $^{238}$U and $^{232}$Th, although the operations have been less effective in the removal of $^{210}$Pb. Purification efforts are ongoing.

In the near future, another organic liquid scintillator detector, SNO+~\cite{snoplus}, will start taking data. It is located in SNOLAB, which is two times deeper underground than LNGS and, due to the lower muon flux, the $^{11}$C rate is expected to be negligible. Additionally, it is three times larger than Borexino, allowing it to accumulate statistics at a higher rate. Assuming the same radioactive levels as Borexino, it should be able to measure the pep neutrino interaction rate with 5\% uncertainty, testing the MSW transition region. A measurement of the solar neutrino flux from the CNO cycle is also expected.

Beyond the measurement of pep and CNO solar neutrinos, it may be possible to measure the lowest energy solar neutrinos: pp neutrinos and the monoenergetic 0.38\,MeV $^7$Be neutrinos (see Figure~\ref{fig:snu_spec_flux}). Due to the intrinsic $^{14}$C background in organic scintillators, which dominates the low energy region of the spectrum, these measurements may require new technologies. Noble liquid detectors like CLEAN~\cite{clean} and XMASS~\cite{xmass} have high scintillation yields and no intrinsic radioactive background, making the measurement possible, although the target masses required are larger than that of the current detectors. Another possibility is an organic liquid scintillator with a delayed tag for neutrino interactions, e.g. LENS~\cite{lens}. The measurement of the pp neutrino flux would be direct evidence for the fundamental reaction that fuels the Sun and will offer a high precision test of MSW-LMA and the SSM, as its rate is predicted with the smallest uncertainty. Measurement of the 0.38\,MeV $^7$Be neutrinos, together with the corresponding 0.86\,MeV line, would provide a SSM-independent relative electron neutrino survival probability measurement at two energies in the vacuum-dominated region.

Although it is hard to predict the solar neutrino performance of a 50 kiloton-scale liquid scintillator detector like LENA~\cite{lena}, the measurement of the flux from the CNO electron-capture neutrino lines at 2.2\,MeV ($^{13}$N) and 2.8\,MeV ($^{15}$O)~\cite{cnocapture}, which fall in an unexplored section of the MSW transition region, can only be possible in such a large detector.

\section{Conclusion}

After more than 40 years, solar neutrinos continue to be an important tool in the study of fundamental physics and stellar astrophysics. In the last five years, Borexino has overcome the backgrounds from natural radioactivity to perform sub-MeV solar neutrino spectroscopy. The measurement of the $^7$Be neutrino flux has confirmed that low energy neutrinos undergo vacuum-dominated oscillations, a key feature of the leading oscillation model, MSW-LMA. Furthermore, Borexino has taken the first step toward the measurement of pep neutrinos and neutrinos from the CNO cycle, opening the door for high precision tests of solar neutrino oscillations and the resolution of outstanding issues in solar astrophysics. With the advent of SNO+, solar neutrinos may become one of the most sensitive probes in the search for new physics and in the understanding of the composition of the solar core.

\bigskip 
\bibliography{basename of .bib file}

\end{document}